\input harvmac
\input epsf

\Title{ DFTUZ /96/02}
{\vbox{\centerline{Bethe ansatz equations for quantum chains }
\vskip2pt\centerline{combining different representations of $su(3)$ }}}
\centerline{J. Abad and M. Rios}
\centerline{Departamento de F\'{\i}sica Te\'{o}rica, Facultad de Ciencias,}
\centerline{Universidad de Zaragoza, 50009 Zaragoza, Spain}
\bigskip
\bigskip
\vskip .3in
\centerline{ \tenbf Abstract}
The general expression for the local matrix of a quantum chain  $L(\theta)$
with the site space in any representation of  $su(3)$ is obtained.  This is
made by generalizing
$L(\theta)$ from the fundamental representation and imposing the fulfilment of
the Yang-Baxter equation. With these operators and using a generalization of
the nested Bethe ansatz, the Bethe equations for a multistate quantum chain
combining two arbitrary representations of $su(3)$ are obtained .
 \vskip .2in
\bigskip
\bigskip
\Date{}

\vfill
\eject
In the study of integrable quantum systems, chains combining two kinds of
spin have aroused great interest lately. The work was pioneered for
$su(2)$ algebra by H. de Vega and F. Woynarovich  \ref\ri{H.J. de Vega and F.
Woynarovich, J. Phys. A25 (1992) 4499.}. In this paper a chain mixing
sites with spin $1/2$ and $1$ and periodic boundary conditions was studied, and
 the generalization to a chain combining spin 1/2 and any other $s$ was
suggested.
Several subsequent works have been published in which the thermodynamic
properties of these systems are studied (\ref\rii{H.J. de Vega, L. Mezincescu
and R.I.
Nepomechie, Phys. Rev. B49 (1994) 13223.}\nref\riii{H.J. de Vega, L.
Mezincescu and R.I. Nepomechie, J. Mod. Phys. B8 (1994)
3473.}\nref\riv{M.J. Martins, J. Phys. A 26 (1993) 7301.}--\ref\rv{S. R.
Aladin and M.J. Martins, J Phys. A 26 (1993) 1529 and J. Phys A 26
(1993) 7287.}).

In this paper, we study an alternating chain the site states of which are a
mixture of any two representations of $su(3)$. We made an initial approach to
this
problem in  a previous paper \ref\rvi{J. Abad and M.  R\'{\i}os, DFTUZ 95/23.
To be published in Phys. Rev. B.},
where we solved an alternating chain mixture of the two fundamental
representations of  $su(3)$ and presented a method, a modification of the
nested
Bethe ansatz (MNBA), needed to find the  Bethe equation (BE)
solutions of the problem. The process was as follows: in first place we sought
the general form of the local operator $L(\theta)$ with its
auxiliary space in the fundamental representation (\ref\vii{M. Jimbo, Commun.
Math. Phys. 102 (1986) 537.}\nref\viii{V.V.
Bazhanov, Phys. Lett B 159 (1985) 321 and Commun. Math. Phys. 113
(1987) 471.}\nref\ix{M. Jimbo, T Miwa and M. Okado, Mod. Phys. Lett.
B1 (1987) 73.}--\ref\x{H.J. de Vega, J. Mod. Phys. A4 (1989) 2371 \semi H.J.
de Vega, Nucl. Phys. B (proc. nad Suppl.) 18A (1990) 229 \semi J. Abad and
 M. R\'{\i}os, Univ. of Zaragoza preprint DFTUZ 94-11 (1994).}) and  the site
space in any representation of $su(3)$. This is done by departing from a
general form  inspired by the local operator $L(\theta)$ with the auxiliary and
site space in the fundamental representation of $su(3)$ and by making that
operator the YBE solution. The operator so obtained has several free
parameters that are coming from the symmetries of the YBE. With this operator
we can form integrable homogeneous chains and find the ansatz equations
with usual nested Bethe ansatz (NBA) (\ref\xi{N. Reshetikhin and P.B. Wiegmann,
Phys. Lett. B 189 (1987) 125.},\ref\xii{V.V. Bazhanov and N.
Reshetikhin, J. Phys. A 23 (1990) 1477.}). In a second step,  alternating
chains are formed by mixing any two representations of $su(3)$ and the
solutions are formed by applying MNBA
ref. \rvi . From the results so obtained we can conjecture the BE for chains
based on the algebra $su(n)$.

We denote a representation by the indices of its associated Dynkin diagram
$(m_1, m_2)$, where $m_1$ and $m_2$ correspond to the $\{3\}$ and
$\{\bar{3}\}$ representations respectively. In the figures a continuous
line was used for the fundamental representation $(1,0)$ and a wavy line for
any other
representation. Thus, the operators $L(\theta)$  are denoted as indicated in
figure~1 and in order to simplify the writing of the formulae, we will
adopt the following identifications: $L(\theta) \equiv L^{(1,0)(1,0)}(\theta)$
and
${L'}(\theta) \equiv L^{(1,0)(m_{1},m_{2})}(\theta)$.

\midinsert
\bigskip
\centerline{\epsfxsize=10cm  \epsfbox{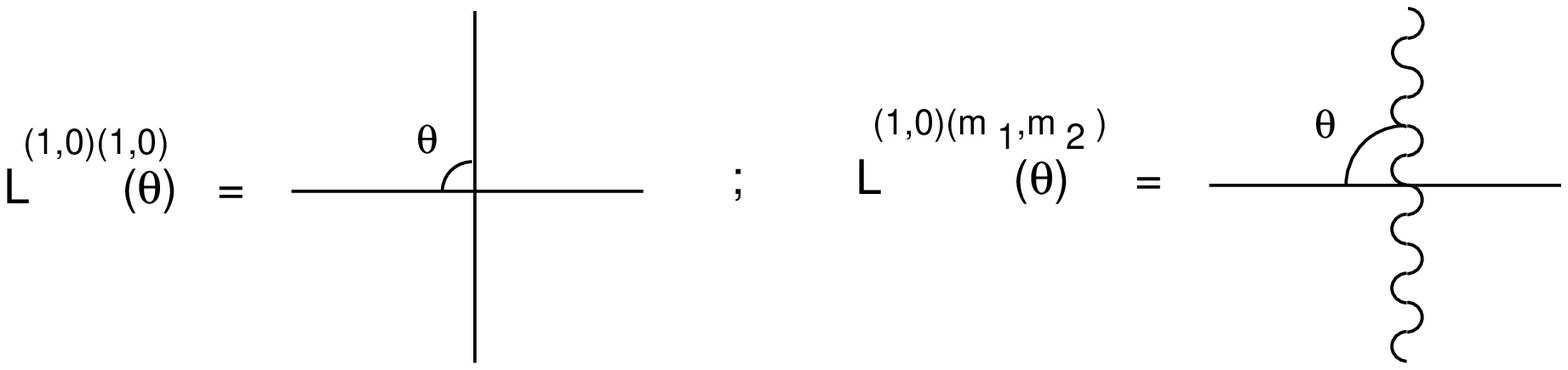}}
\centerline{Figure. 1}
\endinsert
\bigskip

The operator $L(\theta)$ can be written \rvi
\eqn\ei{
L(\theta)=
\pmatrix{
{1 \over 2}(\lambda^{3}q^{-N^{\alpha}}-\lambda^{-3}q^{N^{\alpha}}) & \lambda
{(q^{-1}-q)
\over 2} f_{1} & \lambda^{-1}{(q^{-1}-q) \over 2} [f_{2},f_{1}] \cr
\lambda^{-1}{(q^{-1}-q) \over 2} e_{1} &{1 \over 2}(\lambda^{3}q^{-N^{\beta}}-
\lambda^{-3}q^{N^{\beta}}) & \lambda {(q^{-1}-q)\over 2} f_{2} \cr
\lambda{(q^{-1}-q) \over 2} [e_{1},e_{2}]  & \lambda^{-1}{(q^{-1}-q) \over 2}
e_{2} &
{1 \over 2}(\lambda^{3}q^{-N^{\gamma}}-\lambda^{-3}q^{N^{\gamma}}) \cr
},}
where the parameters  $\lambda$ and $q$ have be taken as the functions of
$\theta$ and $\gamma$
\eqn\eii{\lambda = e^{\theta \over 2}, \qquad q = e^{-\gamma}}
and the $N$ matrices are
\eqna\eai
$$\eqalignno{N^{\alpha}&={2 \over 3}h_{1}+{1 \over 3}h_{2}+{1 \over 3}I,&\eai a
\cr
N^{\beta}&=-{1 \over 3}h_{1}+{1 \over 3}h_{2}+{1 \over 3}I,&\eai b \cr
N^{\gamma}&=-{1 \over 3}h_{1}-{2 \over 3}h_{2}+{1 \over 3}I,&\eai c \cr}
$$
where $\{e_{i}, f_{i}, q^{\pm h_{i}}\}$, $i=1, 2$ the Cartan generators of the
deformed algebra $U_{q}(sl(3))$.

To obtain the operators ${L'}(\lambda) $ with the new parameters given in \eii,
we take \ei\ as a basis and write
\eqn\eiv{
L'(\lambda)=
\pmatrix{
{1 \over 2}(\lambda^{3}q^{-N^{\alpha}}-\lambda^{-3}q^{N^{\alpha}}) & \lambda
 F_{1} & \lambda^{-1} F_{3} \cr
\lambda^{-1} E_{1} &{1 \over 2}(\lambda^{3}q^{-N^{\beta}}-
\lambda^{-3}q^{N^{\beta}}) & \lambda  F_{2} \cr
\lambda E_{3}  & \lambda^{-1} E_{2} &
{1 \over 2}(\lambda^{3}q^{-N^{\gamma}}-\lambda^{-3}q^{N^{\gamma}}) \cr
}.}
where the operators $\{E_{i},F_{i}\}$, $i=1,3$ are  unknown and will be
determined by imposing the YBE
\eqn\ev{R(\lambda/\mu)[L'(\lambda) \otimes L'(\mu)] =
[L'(\mu) \otimes L'(\lambda)]R(\lambda/\mu)
}
\midinsert
\bigskip
\centerline{\epsfxsize=10cm  \epsfbox{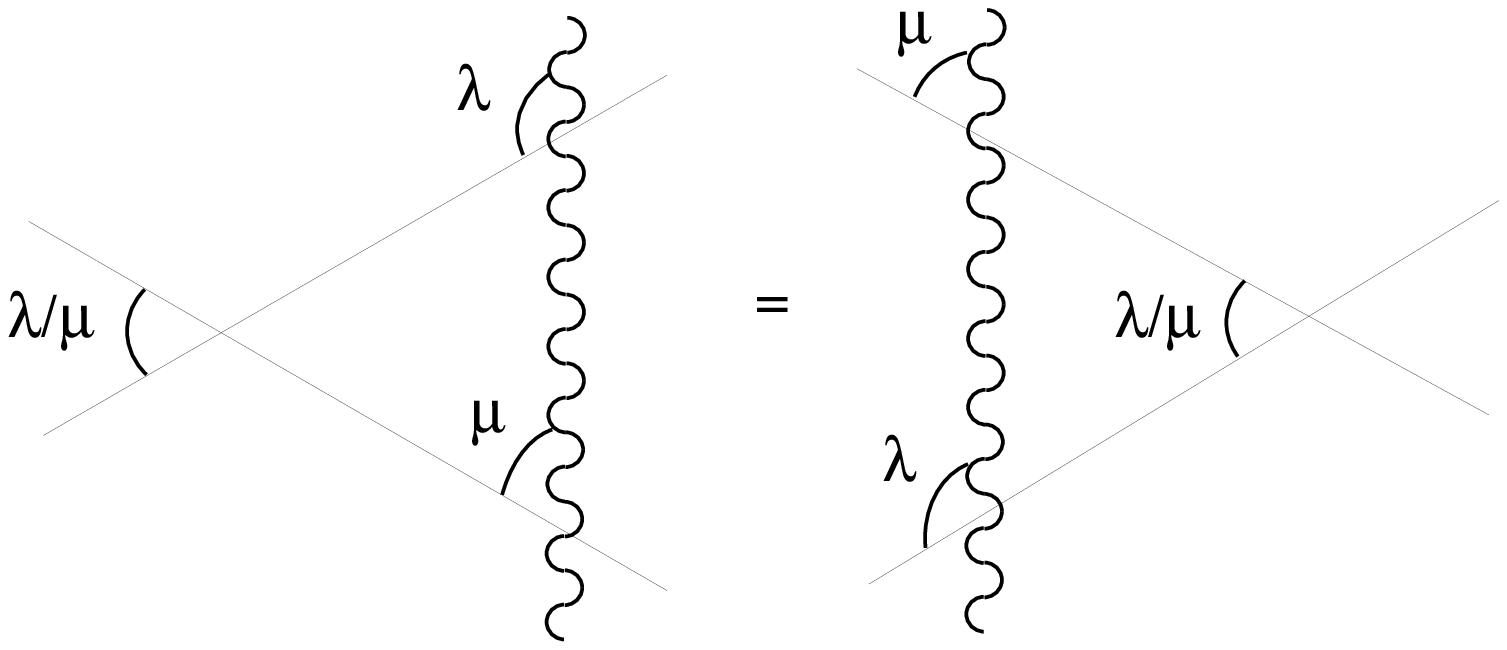}}
\centerline{Fig. 2}
\endinsert
\bigskip
\noindent as shown in figure 2.
The $R_{c, a} ^ {b,d}(\theta)\equiv [ L_{a,b} (\theta)]_{c,d}$ is given \x
\eqn\evi{
R(\lambda,\mu)=\left (\matrix{
a & 0 & 0 & 0 & 0 & 0 & 0 & 0 & 0 \cr
0 & d & 0 & b & 0 & 0 & 0 & 0 & 0 \cr
0 & 0 & c & 0 & 0 & 0 & b & 0 & 0 \cr
0 & b & 0 & c & 0 & 0 & 0 & 0 & 0 \cr
0 & 0 & 0 & 0 & a & 0 & 0 & 0 & 0 \cr
0 & 0 & 0 & 0 & 0 & d & 0 & b & 0 \cr
0 & 0 & b & 0 & 0 & 0 & d & 0 & 0 \cr
0 & 0 & 0 & 0 & 0 & b & 0 & c & 0 \cr
0 & 0 & 0 & 0 & 0 & 0 & 0 & 0 & a \cr
}\right ),
}
with
\eqna\eri
$$\eqalignno{a(\lambda,\mu)&={1 \over 2}(\lambda^{3}\mu^{-3}q^{-1} -
\lambda^{-3}\mu^{3}q) ,&\eri a  \cr
b(\lambda,\mu)&={1 \over 2}(\lambda^{3}\mu^{-3} -
\lambda^{-3}\mu^{3}) ,&\eri b \cr
c(\lambda,\mu)&={1 \over 2}(q^{-1}-q)\lambda\mu^{-1} ,&\eri c \cr
d(\lambda,\mu)&={1 \over 2}(q^{-1}-q)\lambda^{-1}\mu .&\eri d \cr}
$$

The relations obtained are
\eqna\eaii
$$\eqalignno{E_{1}q^{N^{\alpha}} &= q^{-1}q^{N^{\alpha}} E_{1}, &\eaii a  \cr
E_{1}q^{N^{\beta}} &= qq^{N^{\beta}} E_{1}, &\eaii b \cr
F_{1}q^{N^{\alpha}} &= qq^{N^{\alpha}} F_{1}, &\eaii c \cr
F_{1}q^{N^{\beta}} &= q^{-1} q^{N^{\beta}} F_{1}, &\eaii d \cr
E_{2}q^{N^{\alpha}} &= qq^{N^{\alpha}} E_{2}, &\eaii e \cr
E_{2}q^{N^{\beta}} &= q^{-1}q^{N^{\beta}} E_{2}, &\eaii f \cr
F_{2}q^{N^{\alpha}} &= q^{-1}q^{N^{\alpha}} F_{2},&\eaii g \cr
F_{2}q^{N^{\beta}} &= qq^{N^{\beta}} F_{2} ,&\eaii h \cr
[E_{1},F_{1}]&=(q^{-1}-q)\left(q^{N^{\beta}-N^{\alpha}} -
q^{N^{\alpha}-N^{\beta}} \right), &\eaii i \cr
[E_{2},F_{2}]&=(q^{-1}-q)\left(q^{N^{\gamma}-N^{\beta}} -
q^{N^{\beta}-N^{\gamma}} \right), &\eaii j \cr
E_{3} &= {1 \over (q^{-1}-q)}q^{-N^{\beta}}[E_{1},E_{2}] ,&\eaii k \cr
F_{3} &= {1 \over (q^{-1}-q)}q^{N^{\beta}}[F_{2},F_{1}] ,&\eaii l \cr}
$$
and besides, the modified Serre relations
\eqna\eaiii
$$\eqalignno{&q^{-1}E_{1}E_{1}E_{2}-(q+q^{-1})E_{1}E_{2}E_{1} +
qE_{2}E_{1}E_{1} =0, &\eaiii a  \cr
&qE_{2}E_{2}E_{1}-(q+q^{-1})E_{2}E_{1}E_{2} +
q^{-1}E_{1}E_{2}E_{2} = 0, &\eaiii b \cr
&q^{-1}F_{1}F_{1}F_{2}-(q+q^{-1})F_{1}F_{2}F_{1} +
qF_{2}F_{1}F_{1} = 0, &\eaiii c \cr
&qF_{2}F_{2}F_{1}-(q+q^{-1})F_{2}F_{1}F_{2} +
q^{-1}F_{1}F_{2}F_{2} = 0 ,&\eaiii d \cr}
$$
should be verified.

It must be noted that that the relations \eaii\ are the usual ones for the
quantum group
$U_{q}(sl(3))$ while the relations \eaiii\ are not the usual ones for the said
group and
because of this the EYB is not verified if the
generators  $e_i$ and $f_i$, pertaining to deformed algebra, are taken as $E_i$
and $F_i$. This induces us to take
\eqna\eaiv
$$\eqalignno{F_{i}&={1 \over 2}(q^{-1}-q)Z_{i}f_{i}, &\eaiv a  \cr
E_{i}&={1 \over 2}(q^{-1}-q)e_{i}Z_{i}^{-1} ,\qquad \qquad i=1,2 &\eaiv b \cr}
$$
where $e_i$ and $f_i$, $i=1,2$ are the generators of $U_{q}(sl(3))$ in the
representation $(m_1 , m_2)$ and $Z_i$ are two diagonal operators that were
obtained by imposing the verification of the relations \eaii\ and \eaiii\ . In
this way, one obtains the general form of these operators given by
\eqna\eav
$$\eqalignno{Z_{1}&=q^{a_{1}h_{1}-{1 \over 3}h_{2}+a_{3}I} &\eav a  \cr
Z_{2}&=q^{{1 \over 3}h_{1}+(a_{1}+{1 \over 3})h_{2}+b_{3}I} &\eav b \cr}
$$
where the operators $h_i$, $i=1,2$ are the diagonal elements of the algebra
$sl(3)$, and $a_1, a_3$ and $ b_3$ are free parameters that are associated with
the transformations that leave the EYB invariant.

The knowledge of the operator $L' $ permits us to find the ansatz of any
multistate  chain that mixes various representations. For this purpose, the
monodromy operator
corresponding  to the chain to be solved is built; as an example we will use
the
one which alternates the representations $(1, 0)$ and $(m_1, m_2)$
\eqn\evii{
T^{\rm (alt)}_{a,b} (\theta)= L^{(1)}_{a,a_1}(\theta) {L' }
^{(2)}_{a_1,a_2}
(\theta) \ldots
L^{(2N-1)}_{a_{2N-2},a_{2N-1}}(\theta){L' }^{(2N)}_{a_{2N-1,b}}
(\theta) ,
}
that can be represented graphically as shown in figure 3.
\midinsert
\bigskip
\centerline{\epsfxsize=10cm  \epsfbox{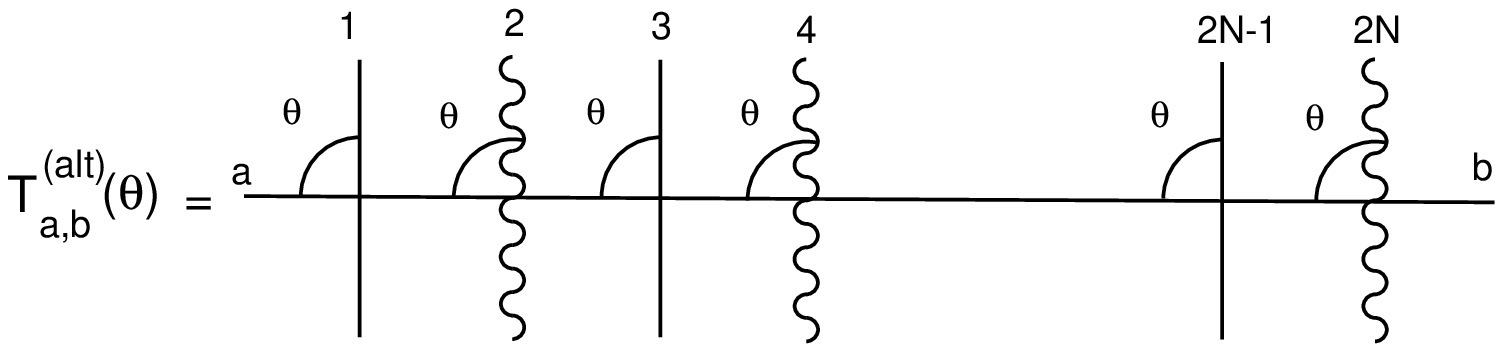}}
\centerline{Figure 3}
\endinsert
\bigskip

Using the MNBA (\rvi, \ref\rxiii{H. J. de Vega and M. Karowski, Nucl. Phys. B
280 (1987) 225.}) the ansatz for the chain can be found. To particularize to
each
case it is necessary to know the action of the diagonal operators $ T^{\rm
alt}_{i, i} $ on the vacuum state if the chain is homogeneous or on the vacuum
subspace if it is an alternating chain \rvi . In both cases, it always
characterized by the highest weight  of the representation. Thus, for the
representation $(m_1, m_2)$ it will be
\eqn\eviii{\Lambda_{h}={2m_{1}+m_{2} \over 3}\alpha_{1} + {m_{1}+2m_{2} \over
3}\alpha_{2} ,
}
where  $\alpha_{1} $ and $\alpha_{2}$ are the simple roots of $su(3)$.

Through \eiv , \eai  and \eviii , together with the commutation rules of
$su(3)$  it was possible to know the action of $L'_{i, i}(\theta)$ on the
highest
weight, obtaining
\eqna\eavi
$$\eqalignno{{L'}_{1,1}(\theta)\Lambda_{h} &=\sinh(
{3 \over 2}\theta+({2 \over 3}m_{1}+{1 \over 3}m_{2}+
{1 \over 3})\gamma)\Lambda_{h}, &\eavi a  \cr
{L'}_{2,2}(\theta)\Lambda_{h} &=\sinh
({3 \over 2}\theta+(-{1 \over 3}m_{1}+{1 \over 3}m_{2}+
{1 \over 3})\gamma)\Lambda_{h}, &\eavi b \cr
{L'}_{3,3}(\theta)\Lambda_{h} &=\sinh
({3 \over 2}\theta+(-{1 \over 3}m_{1}-{2 \over 3}m_{2}+
{1 \over 3})\gamma)\Lambda_{h} .&\eavi c \cr}
$$
It is also applicable for obtaining the action of the operators $L_{i,
i}(\theta)$
on the corresponding  highest weight state taking $m_1= 1$ and $m_2= 0$. In
this way, in the alternate chain that mixes $N$ representations $(1, 0)$ with
$N$
representations $(m_1,  m_2)$, the BE are given by
\eqna\eavii
$$\eqalignno{&[g(\mu_{k})]^{N}[\widetilde{g}_{1}(\mu_{k})]^{N}
=\prod_{j=1 \atop j \neq
k}^{r} {g(\mu_{k}-\mu_{j}) \over g(\mu_{j}-\mu_{k})}\prod_{i=1}^
{s}g(\lambda_{i}-\mu_{k}),&\eavii a \cr
&[\widetilde{g}_{2}(\lambda_{k})]^{N}=\prod_{j=1}^{r}g(\lambda_{k}-\mu_{j})
\prod_{i=1 \atop i \neq
k}^{s} {g(\lambda_{i}-\lambda_{k}) \over g(\lambda_{k}-\lambda_{i})}, &\eavii b
\cr}
$$
where $\mu_i$, $i=1,\cdots, r$ and $\lambda_j$, $j=1,\cdots, s$ the roots of
the ansatz, the function $g$ is,
\eqn\eix{g(\theta)={\sinh({3 \over 2}\theta + \gamma) \over \sinh({3 \over
2}\theta)}}
and $\widetilde{g}_{1}(\theta)$ and $\widetilde{g}_{2}(\theta)$ are obtained
from \eavi\ giving
\eqna\eaviii
$$\eqalignno{\widetilde{g}_{1}(\theta) &={\sinh(
{3 \over 2}\theta+({2 \over 3}m_{1}+{1 \over 3}m_{2}+
{1 \over 3})\gamma) \over \sinh
({3 \over 2}\theta+(-{1 \over 3}m_{1}+{1 \over 3}m_{2}+
{1 \over 3})\gamma)}, &\eaviii a  \cr
\widetilde{g}_{2}(\theta) &={\sinh
({3 \over 2}\theta+(-{1 \over 3}m_{1}-{2 \over 3}m_{2}+
{1 \over 3})\gamma) \over \sinh
({3 \over 2}\theta+(-{1 \over 3}m_{1}+{1 \over 3}m_{2}+
{1 \over 3})\gamma)}. &\eaviii b \cr}
$$

The procedure can be generalized to chains that mix non-fundamental
representations, irrespective of the number of sites and their distribution
in the representations. For this purpose, it is necessary to build the
monodromy matrix
following an analogous process to use in \evii . If we use a dashed line for
the
representation $({m'}_1, {m'}_2)$, the monodromy matrix $T^{\rm (gen)}(\theta)$
can be represented  graphically as shown in figure 4.
\midinsert
\bigskip
\centerline{\epsfxsize=10cm  \epsfbox{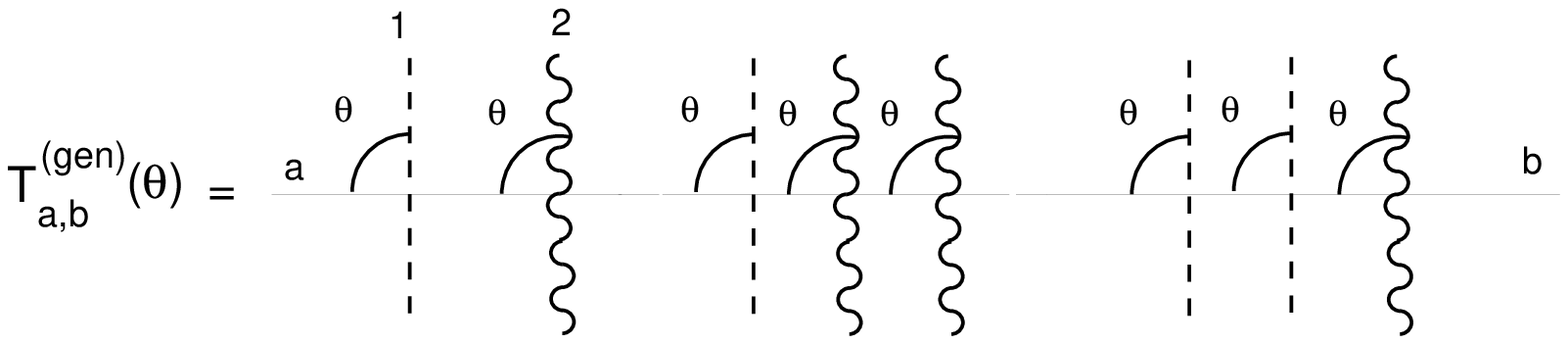}}
\centerline{Figure 4}
\endinsert
\bigskip

The eigenvalues for the local operators on the highest weight states, in
straighforward notation are
\eqna\eaixn
$$\eqalignno{
\bar{l}_{1,1}(\theta) = & \sinh ({3 \over 2} \theta +
	({2 \over 3}m_{1} +{ 1\over 3}m_{2} + {1 \over 3})\gamma)
	&\eaixn a \cr
\bar{l}_{2,2}(\theta) = & \sinh ({3 \over 2} \theta +
	(-{1 \over 3}m_{1} +{ 1\over 3}m_{2} + {1 \over 3})\gamma)
	&\eaixn b \cr
\bar{l}_{3,3}(\theta) = & \sinh ({3 \over 2} \theta +
	(-{1 \over 3}m_{1} -{ 2\over 3}m_{2} + {1 \over 3})\gamma)
	&\eaixn c \cr
\widetilde{l}_{1,1}(\theta) = & \sinh ({3 \over 2} \theta +
	({2 \over 3}m'_{1} +{ 1\over 3}m'_{2} + {1 \over 3})\gamma)
	&\eaixn d \cr
\widetilde{l}_{2,2}(\theta) = & \sinh ({3 \over 2} \theta +
	(-{1 \over 3}m'_{1} +{ 1\over 3}m'_{2} + {1 \over 3})\gamma)
	&\eaixn e \cr
\widetilde{l}_{3,3}(\theta) = & \sinh ({3 \over 2} \theta +
	(-{1 \over 3}m'_{1} -{ 2\over 3}m'_{2} + {1 \over 3})\gamma)
	&\eaixn f \cr}
$$
By calling the number of sites in the representations $({m}_1,
{m}_2)$ and $({m'}_1, {m'}_2)$ $N_1$ and  $N_2$ respectively, we found the
eigenvalue of the transfer matrix for this general chain
\eqn\eaixnn{
\eqalign{
\Delta (\theta)=&[\bar{l}_{1,1}(\theta)]^{N_1}
[\widetilde{l}_{1,1}(\theta)]^{N_2}\prod_{j=1}^{r} g(\mu_j -\theta) +  \cr
&\prod_{j=1}^{r}  g(\theta-\mu_j ) \Bigl[  [\bar{l}_{2, 2}(\theta)]^{N_1}
[\widetilde{l}_{2,2}(\theta)]^{N_2}\prod_{i=1}^{s} g(\lambda_i -\theta) +  \cr
& [\bar{l}_{3, 3}(\theta)]^{N_1}
[\widetilde{l}_{3,3}(\theta)]^{N_2}\prod_{l=1}^{r}{ 1\over {g(\theta-\mu_l)}}
\prod_{i=1}^{s}g( \theta-\lambda_i) \Bigr]  \cr
}}
and the BE are

\eqna\eaix
$$\eqalignno{[\bar{g}_{1}(\mu_{k})]^{{N}_1}[\widetilde{g}_{1}(\mu_{k})]^{{N}_2}
&=\prod_{j=1 \atop j \neq
k}^{r} {g(\mu_{k}-\mu_{j}) \over g(\mu_{j}-\mu_{k})}\prod_{i=1}^
{s}g(\lambda_{i}-\mu_{k}) &\eaix a \cr
[\bar{g}_{2}(\lambda_{k})]^{{N}_1}[\widetilde{g}_{2}(\lambda_{k})]^{{N}_2}&=
\prod_{j=1}^{r}g(\lambda_{k}-\mu_{j})
\prod_{i=1 \atop i \neq
k}^{s} {g(\lambda_{i}-\lambda_{k}) \over g(\lambda_{k}-\lambda_{i})}
&\eaix b \cr}
$$
where $\widetilde{g}_{1}$ and $\widetilde{g}_{2}$ are given in \eavii{a,b}\ and
 $\bar{g}_{1}$ and $\bar{g}_{2}$ are the same as the previous ones but
$(m_{1},m_{2})$ is replaced by $(m'_{1},m'_{2})$.

In the light of this, the generalization for the case of
mixed chains with more than two different representations seems simple,although
the physical
models that they represent will be less local and the interaction more complex.

In a non-homogeneous chain  combining different representations of $su(n)$,
each  representation introduces $(n-1)$
functions ( that we call source functions). Each solution will have $(n-1)$
sets of equations (the same number of dots in its Dynkin diagram). The first
member of the equations will be a product of the respective source functions
powered to the number of sites of each representation and the second a product
of source functions similar to \eaix\ .

\noindent{ \tenbf  Acknowledgements}

We are grateful to professor H. de Vega for his very useful discussions and
remarks. A
careful reading of the manuscript by professor J. Sesma is also acknowledged.
This work was partially supported by the Direcci\'{o}n General de
Investigaci\'{o}n Cient\'{\i}fica y T\'{e}cnica, Grant No PB93-0302 and
AEN94-0218

\listrefs
\end{document}